\documentclass[twocolumn,aps,pra,showpacs,superscriptaddress]{revtex4}
\usepackage[usenames,dvipsnames]{color}

\usepackage{graphicx}
\usepackage{amsmath,amssymb}
\usepackage{color}
\usepackage{graphics}
\usepackage{epsfig}
\usepackage{hyperref}
\hypersetup{breaklinks}

\newcommand{\be}{\begin{equation}}
\newcommand{\ee}{\end{equation}}
\newcommand{\eea}{\end{eqnarray}}
\newcommand{\bea}{\begin{eqnarray}}

\newcommand{\mean}[1]{\ensuremath{\langle{#1}\rangle}}

\newcommand{\eins}{\openone}
\newcommand{\qed}{\ensuremath{\hfill \Box}}

\newcommand{\af}{\ensuremath{\mathfrak{a}}}

\newcommand{\bfrak}{\ensuremath{\mathfrak{b}}}

\newcommand{\XX}{\ensuremath{\mathcal{X}}}

\newcommand{\ZZ}{\ensuremath{\mathcal{Z}}}

\newcommand{\ketbra}[1]{\ensuremath{| #1 \rangle \!\langle #1 |}}

\newcommand{\ket}[1]{\ensuremath{|#1\rangle}}

\newcommand{\braket}[2]{\ensuremath{\langle #1|#2\rangle}}

\newcommand{\kommentar}[1]{}
\newcommand{\trace}{{\rm tr}}

\renewcommand{\vr}{\ensuremath{\varrho}}

\newcommand{\forget}[1]{}

\begin{document}


\title{Bounding the quantum dimension with contextuality}



\author{Otfried G{\"u}hne}
 \affiliation{Naturwissenschaftlich-Technische Fakult\"at,
Universit\"at Siegen,
Walter-Flex-Str.~3,
D-57068 Siegen}

\author{Costantino Budroni}
\affiliation{Naturwissenschaftlich-Technische Fakult\"at,
Universit\"at Siegen,
Walter-Flex-Str.~3,
D-57068 Siegen}
\author{Ad\'{a}n Cabello}
\affiliation{Departamento de F\'{\i}sica Aplicada II,
 Universidad de Sevilla, E-41012 Sevilla, Spain}

\author{Matthias Kleinmann}
\affiliation{Naturwissenschaftlich-Technische Fakult\"at,
Universit\"at Siegen,
Walter-Flex-Str.~3,
D-57068 Siegen}
\author{Jan-{\AA}ke Larsson}
\affiliation{Institutionen f\"or Systemteknik och Matematiska Institutionen, Link\"opings Universitet, SE-581 83 Link\"oping, Sweden}


\date{\today}


\begin{abstract}
We show that the phenomenon of quantum contextuality can be used
to certify lower bounds on the dimension accessed by the
measurement devices. To prove this, we derive bounds for different 
dimensions and scenarios of the simplest noncontextuality inequalities. 
The resulting dimension witnesses work independently of the prepared
quantum state. Our constructions are robust against noise and
imperfections, and we show that a recent experiment can be viewed
as an implementation of a state-independent quantum dimension
witness.
\end{abstract}




\maketitle


\section{Introduction}
The recent progress in the experimental control and manipulation
of physical systems at the quantum level opens new possibilities
(e.g., quantum communication, computation, and simulation), but,
at the same time, demands the development of novel theoretical
tools of analysis. There are already tools which allow us to
recognize quantum entanglement and certify the usefulness of
quantum states for quantum information processing tasks \cite{HHH96, GT09}.
However, on a more fundamental level, there are still several problems
which have to be addressed. For example, how can one efficiently test
whether measurements actually access all the desired energy levels
of an ion? How to certify that the different paths of photons
in an interferometer can be used to simulate a given multi-dimensional
quantum system? Similar questions arise in the analysis of experiments
with orbital angular momentum, where high-dimensional entanglement
can be produced \cite{MTT07,dada11}, or in experiments with
electron spins at nitrogen-vacancy centers in diamond, where the
quantumness of the measurements should be certified
\cite{NMRHWYJGJW08}.

The challenge is to provide lower bounds on the {\em dimension}
of a quantum system only from the statistics of measurements
performed on it. More precisely, one certifies lower bounds on 
the dimension of the underlying Hilbert space, where the 
measurement operators act on. Such bounds can  be viewed
as lower bounds on the complexity and the number of levels 
accessed by the measurement devices: If the measurement operators 
act non-trivially only on a small subspace, then all measurements 
results can be modeled by using a low-dimensional quantum system 
only. Note that this is not directly related to the rank of a density
matrix. In fact, a pure quantum state acting on a one-dimensional subspace
only can still give rise to measurement results, which can only be explained 
assuming a higher-dimensional Hilbert space.

The problem of estimating the Hilbert space dimension has been considered in 
different scenarios, and slightly different notions of dimension were involved.
Brunner and coworkers introduced the concept of quantum 
``dimension witnesses'' by providing lower bounds
on the dimension of composite systems from the violation of Bell
inequalities \cite{BPAGMS08, VP09}. The nonlocal properties of the correlations
produced are clearly the resource used for this task. 
As a consequence, even if the experimenter is able to access and manipulate many levels of her systems locally, but she is not able to entangle those levels, the above test fails to certify such a dimension.
Such a task can therefore be interpreted as a test of the type of entanglement and correlations produced, namely, how many levels or degrees of freedom the experimenter is able to entangle.

In a complementary scenario, several different states of a single
particle are prepared and different measurements are carried out
\cite{WCD08, GBHA10, BNV}. This approach has also recently been implemented
using photons \cite{HGMBAT12, ABCB12}. In this situation, the dimension of the 
system can be interpreted as the dimension of the set of states the 
experimenter is able to prepare.

As a third possibility, also the continuous time evolution can be
used to bound the dimension of a quantum system \cite{WP09}. In this case,
the relevant notion of dimension is that of the set of states generated by
the dynamical evolution of the system.

In this paper we focus on sequential measurements on a single system, 
a type of measurements used in tests of quantum contextuality, and we 
show how they can be  used for bounding the dimension of quantum 
systems. Quantum contextuality is a genuine quantum effect leading 
to the Kochen-Specker theorem, which states that quantum mechanics is in
contradiction to non-contextual hidden variable (NCHV) models
\cite{Specker60, gleason, Bell66, KS67, LSW11}. In fact, already in the first
formulation of the theorem the dimension plays a central role \cite{Specker60}.

We derive bounds for the several important noncontextuality (NC) inequalities
for different dimensions and scenarios. The experimental violation of these
bounds automatically provides a lower bound on the dimension of the system,
showing that NC inequalities can indeed be used as dimension witnesses.
Remarkably, contextuality can be used as a resource for bounding the
dimension of quantum systems in a state-independent way. 

This illustrates clearly the difference with the existing schemes: 
Dimension witnesses derived according to Refs.~\cite{GBHA10, BNV} 
certify the minimum classical or quantum dimension spanned by a set 
of preparations. They distinguish between classical and quantum 
dimension $d$, but, in general, not between quantum dimension $d$ 
and classical dimension $d+1$. They require at least $d+1$ preparations 
to certify a dimension $d$.
On the other side, dimension witnesses based on Bell's theorem or
contextuality certify the minimum quantum dimension accessed by 
the measurement devices acting on a system prepared in the a single
state. Contrary to the Bell scenario \cite{BPAGMS08, VP09}, in our approach
the initial state and its nonlocal properties play no role
and the result of our test can directly be interpreted as the minimal
number of levels accessed and manipulated by the measurement apparatus.

The paper is organized as follows. In Sec. I, we discuss the case of state-dependent
noncontextuality inequalities, specifically, Klyachko, Can, Binicio\u{g}lu,
and Shumovsky (KCBS) inequality \cite{KCBS08}. In Sec. II, we discuss what happens 
when the sequences of measurements contain non-compatible measurements. In Sec. IV and Sec. V,
we apply the same analysis to the case of state-independent noncontextuality 
inequalities, specifically, the Peres-Mermin (PM) inequality \cite{Peres90,Mermin90,Cabello08}. In Sec. VI,
we discuss the case of imperfect measurements, then in Sec. VII we show how a recent 
experimental test of contextuality can be viewed as an implementation of our dimension witness.


\section{The KCBS inequality}
We first turn to the state-dependent case.
The simplest system showing quantum contextuality
is a quantum system of
dimension three \cite{Specker60}. The simplest NC inequality
in three dimensions is the one introduced by Klyachko, Can, Binicio\u{g}lu,
and Shumovsky (KCBS) \cite{KCBS08}. For that, one considers
\begin{equation}
\mean{\chi_{\rm KCBS}} =
\mean{AB} + \mean{BC} + \mean{CD}
+ \mean{DE} + \mean{EA},
\end{equation}
where $A, B, C, D,$ and $E$ are measurements with outcomes $-1$ and $1$,
and the measurements in the same mean value $\mean{\ldots}$ are compatible
\cite{Peres95}, i.e., are represented in quantum mechanics by commuting
operators. The mean value itself is defined via a sequential measurement: For
determining $\mean{AB}$, one first measures $A$ and then $B$ on the same
system, multiplies the two results, and finally averages over many
repetitions of the experiment.

The KCBS inequality states that
\begin{equation}
\mean{\chi_{\rm KCBS}} \stackrel{\rm NCHV}{\geq} -3,
\label{kcbsa}
\end{equation}
where the notation ``$\stackrel{\rm NCHV}{\geq} -3$'' indicates that $-3$ is the minimum
value for any NCHV theory. Here, noncontextuality means that the theory assigns
to any observable (say, $B$) a value independent of which other
compatible observable (here, $A$ or $C$) is measured jointly with it.

In quantum mechanics, a value of $\mean{\chi_{\rm KCBS}}=5-4\sqrt{5} \approx -3.94$
can be reached on a three-dimensional system, if the observables and the initial
state are appropriately chosen. This quantum violation of the NCHV bound
does not increase in higher-dimensional systems \cite{LSW11, AQBTC12}, and
the violation of the KCBS inequality has been observed in recent
experiments with photons \cite{LLSLRWZ11, AACB12}.

Given the fact that quantum contextuality requires a three-dimensional
Hilbert space, it is natural to ask whether a violation of Eq.~(\ref{kcbsa})
implies already that the system is not two-dimensional. The following
observation shows that this is the case:


{\bf Observation 1.} Consider the KCBS inequality where the measurements
act on a two-dimensional quantum system and are commuting,
i.e., $[A,B]=[B,C]=[C,D]=[D,E]=[E,A]=0.$
Then, the classical bound holds:
\begin{equation}
\mean{\chi_{\rm KCBS}} \stackrel{\rm 2D, com.}{\geq} -3.
\label{kcbs}
\end{equation}


{\it Proof of Observation~1.} First, if two observables
$A$ and $B$ are compatible, then $|\langle A\rangle\pm\langle AB\rangle|\le1\pm\langle B\rangle$.
This follows from the fact that $A$ and $B$ have common eigenspaces
and the relation holds separately on each eigenspace.
Second, in two dimensions, if $[A,B]=0=[B,C]$, then either
$B=\pm \eins$ or $[A,C]=0$. The reason is that, if $B$ is not the
identity, then it has two one-dimensional eigenspaces. These are shared
with $A$ and $C$, so $A$ and $C$ must be simultaneously
diagonalizable.

Considering the KCBS operator $\chi_{\rm KCBS},$ the claim is trivial
if $A,\ldots, E$ are all compatible, because then the relation holds
separately on each eigenspace. It is only possible that not
all of them commute if there are two groups in the sequence $\{A,B,C,D,E\}$ 
of operators separated by identity operators. Without loss of generality, 
we assume that the groups of commuting operators are $\{E,A\}$ and $\{C\}$ 
so that $B= b\eins =\pm \eins$ and $D=d \eins = \pm \eins$. This gives
\begin{align}
\mean{\chi_{\rm KCBS}}& = b\langle A\rangle+ b\langle C\rangle+ d\langle C\rangle+
    d\big(\langle E\rangle+d \langle EA\rangle\big)
\nonumber
\\
    & \geq b\langle A\rangle+ b\langle C\rangle+ d\langle C\rangle
    -1-d\langle A\rangle
\nonumber
\\
    &= (b-d)\langle A\rangle+ (b+d)\langle C\rangle -1 \geq-3
\end{align}
and proves the claim. In this argumentation, setting observables
proportional to the identity does not change the threshold, but in
general it is important to consider this case, as this often
results in higher values.
\qed

It should be added that Observation~1 can also be proved using
a different strategy: Given two observables on a two-dimensional
system, one can directly see that if they commute, then either
one of them is proportional to the identity, or their product
is proportional to the identity. In both cases, one has a classical
assignment for some terms in the KCBS inequality and then one can
check by exhaustive search that the classical bound holds. Details
are given in the Appendix~A1.

Furthermore, Observation~1 can be extended to generalizations of the
KCBS inequality with more than five observables \cite{AQBTC12}: For
that, one considers
\be
\mean{\chi_{N}} = \sum_{i=1}^{N-1} \mean{A_i A_{i+1}} + s\mean{A_N A_1},
\ee
where $s=+1$ if $N$ is odd and $s=-1$ if $N$ is even. For this expression,
the classical bound for NCHV theories is given by $\mean{\chi_{N}} \geq -(N-2).$
In fact, the experiment in Ref.~\cite{LLSLRWZ11} can also be viewed as measurement
of $\mean{\chi_{6}}.$

The discussion of the possible mean values $\mean{\chi_{N}}$ in quantum
mechanics differs for even and odd $N$.
If $N$ is odd, the maximal possible quantum mechanical value is
$\mean{\chi_{N}} = \Omega_N \equiv -[3N\cos(\pi/N)-N]/[1+\cos(\pi/N)]$ and
this value can already be attained in a three-dimensional system
\cite{LSW11, AQBTC12}. The proof of Observation~1 can be generalized in
this case, implying that for two-dimensional systems the classical bound
$\mean{\chi_{N}} \geq - (N-2)$ holds. So, for odd $N$, the generalized KCBS
inequalities can be used for testing the quantum dimension.

If $N$ is even, the scenario becomes richer: First,
quantum mechanics allows to obtain values of
$\mean{\chi_{N}} = \Omega_N \equiv - N \cos(\pi/N),$
but this time this value requires a four-dimensional system \cite{AQBTC12}.
For two-dimensional quantum systems, the classical bound
$\mean{\chi_{N}} \geq - (N-2)$ holds. For three-dimensional systems, one can
show that if the observables $A_i$ in a joint context are different
($A_i \neq \pm A_{i+1}$) and not proportional to the identity, then still
the classical bound holds (for details see Appendix A2). However, if
two observables are the same, e.g. $A_1 = -A_2$, then $\mean{A_1 A_2}=-1$
and $\mean{\chi_{N}}=-1+\mean{\chi_{N-1}}.$ In summary,
for even $N$, we have the following hierarchy of bounds
\be
\mean{\chi_{N}} \stackrel{\rm 2D, com.}{\geq} - (N-2)
\stackrel{\rm 3D, com.}{\geq}
-1 + \Omega_{N-1}
\stackrel{\rm 4D, com.}{\geq}
\Omega_{N}.
\label{kcbsneven}
\ee
Here, the notation $\stackrel{\rm 2D, com.}{\geq}$ etc.~means that 
this bound holds for commuting observables in two dimensions. All 
these bounds are sharp. This shows that extended KCBS inequalities 
are even more sensitive to the dimension than the original inequality.

\section{The KCBS inequality with incompatible observables}
In order to apply Observation~1
the observables must be compatible. Since this condition is not easy
to guarantee in experiments \cite{GKCLKZGR10}, we should
ask whether it is possible to obtain a two-dimensional bound for the
KCBS inequality when the observables are not necessarily compatible.
We can state:

{\bf Observation~2. }If the observables $A,\ldots, E$ are dichotomic
observables but not necessarily commuting, then, for any two-dimensional
quantum system,
\be
\mean{\chi_{\rm KCBS}} \stackrel{\rm 2D}{\geq} -\frac{5}{4}(1+\sqrt{5}) \approx -4.04.
\label{prop4eq1}
\ee
This bound is sharp and can be attained for suitably chosen measurements.

The strategy of proving this bound is the following: If the
observables are not proportional to the identity, one can
write $A=\ketbra{A^+}-\ketbra{A^-}$
and $B=\ketbra{B^+}-\ketbra{B^-}$, and express $\ketbra{A^+}$
and $\ketbra{B^+}$ in terms of their Bloch vectors
$\ket{\af}$ and $\ket{\bfrak}.$ Then, one finds that
\be
\mean{AB}=2 |\braket{A^+}{B^+}|^2-1 = \braket{\af}{\bfrak}.
\label{nicetrick1}
\ee
This property holds for all projective measurements on two-dimensional
systems and is, together with a generalization below [see Eq.~(\ref{trick2})] 
a key idea for deriving dimension witnesses. Note that it implies
that the sequential mean value $\mean{AB}$ is
independent of the initial quantum state and also of the
temporal order of the measurements \cite{Fritz10}. Eq.~(\ref{nicetrick1}) allows us
to transform the KCBS inequality into a geometric inequality for
three-dimensional Bloch vectors. Additional details of the proof are
given in Appendix~A3.

Observation~2 shows that the bound for NCHV theories can be
violated already by two-dimensional systems, if the observables
are incompatible. This demonstrates that experiments, which aim at
a violation of Eq.~(\ref{kcbsa}) also have to test the compatibility
of the measured observables, otherwise the violation can be explained
without contextuality.

It must be added that Observation~2 cannot be used to witness the quantum
dimension, since one can show that Eq.~(\ref{prop4eq1}) holds for all
dimensions \cite{BMKG13}. As we see below,
this difficulty can be surmounted by considering NC inequalities in 
which quantum mechanics reaches the algebraic maximum.


\section{The Peres-Mermin inequality}In order to derive the state-independent
quantum dimension witnesses,
let us consider the sequential mean value \cite{Cabello08},
\begin{align}
\mean{\chi_{\rm PM}} = & \mean{A B C}+ \mean{b c a} +
\mean{\gamma \alpha \beta} + \mean{A \alpha a} + \mean{b B \beta}
\nonumber \\
& - \mean{\gamma c C},
\label{pmoperator}
\end{align}
where the measurements in each of the six sequences are compatible. Then, for NCHV theories
the bound
\begin{equation}
 \mean{\chi_{\rm PM}} \stackrel{\rm NCHV}{\leq} 4
 \label{pmineq}
\end{equation}
holds. In a four-dimensional quantum system, however,
one can take the following square of observables,
known as the Peres-Mermin square \cite{Peres90,Mermin90}
\begin{equation}
\begin{array}{ccc}
A=\sigma_z \otimes \openone, &\;\;\;\;
B= \openone \otimes \sigma_z, &\;\;\;\;
C= \sigma_z \otimes \sigma_z,\\
a= \openone \otimes \sigma_x, &\;\;\;\;
b= \sigma_x \otimes \openone, &\;\;\;\;
c= \sigma_x \otimes \sigma_x,\\
\alpha = \sigma_z\otimes\sigma_x, &\;\;\;\;
\beta = \sigma_x \otimes\sigma_z, &\;\;\;\;
\gamma = \sigma_y \otimes\sigma_y.
\end{array}
\label{mpsquare}
\end{equation}
These observables lead for any quantum state to a value of
$\mean{\chi_{\rm PM}} = 6$, demonstrating state-independent
contextuality. The quantum violation has been observed
in several recent experiments \cite{KZGKGCBR09,ARBC09,MRCL09}.
Note that the sequences in
Eq.~(\ref{pmoperator}) are defined such that each observable
occurs either always in the first or always in the second or
always in the third place of a measurement
a sequence. This difference to the standard version
does not matter at this point (since the observables in any row or
column commute), but it will become important below.

The PM inequality is of special interest for our program since it
is violated up to the algebraic maximum with four-dimensional
quantum systems and the violation is state-independent. Therefore,
this inequality is a good candidate for dimension witnesses without
assumptions on the measurements. First, we can state:


{\bf Observation~3.} If the measurements in the PM inequality are
dichotomic observables on a two-dimensional quantum system and if
the measurements in each mean value are commuting, then one cannot
violate the classical bound,
\be
\mean{\chi_{\rm PM}} \stackrel{\rm 2D,\,com.}{\leq} 4.
\ee
If one considers the same situation on a three-dimensional system,
then the violation is bounded by
\be
\mean{\chi_{\rm PM}} \stackrel{\rm 3D,\,com.}{\leq} 4(\sqrt{5}-1) \approx 4.94.
\label{mpdgleich3}
\ee
These bounds are sharp.

The idea for proving this statement is the following: If one
considers the three commuting observables in each mean value
and assumes that they act on a three-dimensional system, then
three cases are possible: (a) one of the three observables is
proportional to the identity, or (b) the product of two observables
is proportional to the identity, or (c) the product of all three
observables is proportional to the identity. One can directly show
that if case (c) occurs in some mean value, then the classical
bound $\mean{\chi_{\rm PM}} \leq 4$ holds. For the cases (a) and
(b), one can simplify the inequality and finds that it always
reduces to a KCBS-type inequality, for which we discussed already
the maximal quantum values in different dimensions
[see Eq.~\ref{kcbsneven}]. Details are given in Appendix~A4.


\section{The PM inequality with incompatible observables}Let
us now discuss the PM inequality, where the observables are
not necessarily compatible. Our results allow us to obtain directly a bound:

{\bf Observation~4.} Consider the PM operator in
Eq.~(\ref{pmoperator}), where the measurements are
not necessarily commuting projective measurements
on a two-dimensional system. Then we have
\be
\mean{{\chi}_{\rm PM}} \stackrel{\rm 2D}{\leq} 3\sqrt{3} \approx 5.20.
\label{pmbound2dnc}
\ee
{\it Proof.}
One can directly calculate as in the proof of Observation~2 that
for sequences of three measurements on a two-dimensional system
\be
\mean{ABC}= \mean{A}\mean{BC}
\label{trick2}
\ee
holds. Here, $\mean{A} = {\rm tr}(\vr A)$ is the usual expectation
value, and $\mean{BC}$ is the state-independent sequential expectation
value given in Eq.~(\ref{nicetrick1}). With this, we can write:
\begin{align}
\mean{{\chi}_{\rm PM}} = &
\mean{A}(\mean{B C} + \mean{\alpha a})
+
\mean{b}( \mean{c a} +\mean{B \beta})
\nonumber
\\
&+
\mean{\gamma}(\mean{\alpha \beta} - \mean{ c C}).
\end{align}
Clearly, this is maximal for some combination of $\mean{A}=\pm 1$,
$\mean{b}=\pm 1$, and $\mean{\gamma}=\pm 1$. But for any of these
choices, we arrive at an inequality that is discussed
in Lemma~7 in Appendix A3. Note that due to Eq.~(\ref{trick2}) the order
of the measurements matters in the definition of $\mean{{\chi}_{\rm PM}}$
in Eq.~(\ref{pmoperator}). This motivates our choice; in fact, for some
other orders (e.g., $\mean{{\tilde\chi}_{\rm PM}} = \mean{A B C}+ \mean{b c a} +
\mean{\beta \gamma \alpha } + \mean{A \alpha a} + \mean{\beta b B }
- \mean{\gamma c C}$) Eq.~(\ref{pmbound2dnc}) does not hold, and one can reach
$\mean{{\tilde\chi}_{\rm PM}} = 1+\sqrt{9+6\sqrt{3}} \approx 5.404$.
\qed

The question arises whether a high violation of the PM inequality
also implies that the system cannot be three-dimensional and whether
a similar bound as Eq.~(\ref{pmbound2dnc}) can be derived. While
the computation of a bound is not straightforward, a simple argument
shows already that measurements on a three-dimensional systems cannot
reach the algebraic maximum $\mean{{\chi}_{\rm PM}}=6$ for any quantum
state: Reaching the algebraic maximum implies that $\mean{ABC}=1.$ This
implies that the value of $C$ is predetermined by the values of $A$ and $B$ and
the value $A$ of determines the product $BC$. As this holds for any quantum
state, it directly follows that $A,B,C$ (and all the other observables in
the PM square) are diagonal in the same basis and commute, so the bound in
Observation 3 holds. From continuity arguments it follows that there must 
be a finite gap between the maximal value of $\mean{{\chi}_{\rm PM}}$ in 
three dimensions and the algebraic maximum.


\section{Imperfect measurements} 
In actual experimental implementations the measurements may not 
be perfectly projective. It is therefore important to discuss the 
robustness of our method against imperfections.
 
Notice that, since we are considering sequential measurements, 
another possibility for maximal violation of the above inequalities 
is the use of a classical device with memory, able to keep 
track of the measurement performed and adjust the outcomes of 
the subsequent measurements accordingly in order to obtain perfect correlations or anti-correlations. However, as proved in 
Ref.~\cite{Fritz10} and also discussed in Ref.~\cite{BMKG13}, 
such a classical device cannot be simulated in  quantum mechanics 
via projective measurements, more general positive operator valued 
measures (POVMs) are necessary. 

We therefore limit our analysis to some physically motivated noise 
models. A noisy projective measurement $A$ may be modelled by a 
POVM with two effects of the type $E^+=(1-p) \eins/2+ p \ketbra{A^+}$ 
and $E^-=(1-p)\eins/2 + p\ketbra{A^-}$. Then, the probabilities of the
POVM can be interpreted as coming from the following procedure:
With a probability of $p$ one performs the projective measurement
and with a probability of $(1-p)$ one assigns a random outcome.
For this measurement model, one can show that Observation~4
is still valid. Details and a more general POVM are discussed
in Appendix A5. We add that the proof strongly depends on the
chosen measurement order in $\mean{{\chi}_{\rm PM}}$ and that
in any case assumptions about the measurement are made, so the
dimension witnesses are not completely independent of the
measurement device.

The above discussion shows that it is extremely important to test the extent to which the measurements are projective and whether they are
compatible. This can be achieved by performing additional tests.
For instance, one can measure observable $A$
several times in a sequence $\mean{A A A}$ to test whether
the measurement is indeed projective. In addition, one may measure
the sequence $\mean{ABA}$ and compare the results of the two measurements
of $A$, to test whether $A$ and $B$ are compatible. For NC inequalities
it is known how this information can be used to derive correction terms
for the thresholds \cite{GKCLKZGR10}, and similar methods can also be
applied here.


\section{Experimental results}
To stress the experimental
relevance of our findings, let
us discuss a recent ion-trap experiment \cite{KZGKGCBR09}. There,
the PM inequality has been measured with the aim to demonstrate
state-independent contextuality. For our purpose, it is important that
in this experiment also all permutations of the terms in the PM inequality
have been measured. This allows also to evaluate our $\mean{{\chi}_{\rm PM}}$
with the order given in Eq.~(\ref{pmoperator}).
Experimentally, a value $\mean{{\chi}_{\rm PM}}=5.36 \pm 0.05$
has been found. In view of Observation 3, this shows that the data
cannot be explained by commuting projective measurements on a
three-dimensional system. Furthermore,
Observation 4 and the discussion above prove that, even if the
measurements are noisy and noncommuting, the data cannot come
from a two-dimensional quantum system.


\section{Generalizations}
Generalizations of our results
to other inequalities are straightforward:
Consider a general noncontextuality inequality invoking measurement
sequences of length two and three. For estimating the maximal value for
two-dimensional systems
(as in Observations 2 and 4) one transforms all sequential measurements
via Eqs.~(\ref{nicetrick1}) and (\ref{trick2}) into expressions with
three-dimensional Bloch-vectors, which can be estimated. Also noise
robustness for the discussed noise model can be proven, as this follows
also from the properties of the Bloch vectors (cf.~Proposition 12 in
the Appendix). In addition, if a statement as in Observation 3 is desired,
one can use the same ideas as the ones presented here, since they rely on
general properties of commuting observables in three-dimensional space.
Consequently, our methods allow to transform most of the known state-independent
NC inequalities (for instance, the ones presented in
Refs.~\cite{Cabello08, adan10, yo12, kl12})
into witnesses for the quantum dimension.

\section{Discussion and conclusion}
We have shown that the two main noncontextuality inequalities - the KCBS inequality 
(Observation 1) and the Peres-Mermin inequality (Observation 3 and 4) - can be used as
dimension witnesses. In particular, Observation 4 shows that the the 
Peres-Mermin inequality can be 
used to certify the dimension of a Hilbert space independently of the state preparation and in a 
noise robust way. Our methods allow the application of other inequalities, showing that 
contextuality
can  be used as a resource for dimension tests of quantum
systems. Our tests are state-independent, in contrast to the existing
tests. This can be advantageous in experimental implementations, moreover
it shows that one can bound the dimension of quantum systems without
using about the properties of the quantum state.
We hope that our results stimulate further research to answer a central
open question: For which tasks in quantum information processing is
quantum contextuality a useful resource?


\begin{acknowledgments}
We thank Tobias Moroder and Christian Roos for discussions.
This work was supported by the BMBF (Chist-Era network QUASAR),
the EU (Marie Curie CIG 293992/ENFOQI), the FQXi Fund (Silicon 
Valley Community Foundation), the DFG,
and the Project No.\ FIS2011-29400 (Spain).
\end{acknowledgments}

\appendix
\section*{Appendix}

\subsection*{A1: Alternative proof of Observation~1}


For an alternative proof of Observation 1, we need the following Lemma:

{\bf Lemma~5. }If two dichotomic measurements on a two-dimensional
quantum system commute $[A_i, A_{i+1}]=0,$ then either
\begin{itemize}
\item[(a)] one of the observables is proportional to the identity,
$A_i = \pm \eins$ or $A_{i+1} = \pm \eins$ or,
\item[(b)] the product of the two observables is proportional to the identity,
$A_i A_{i+1} = \pm \eins$.
\end{itemize}


{\it Proof of Lemma~5.} This fact can easily be checked: the observables
$A_i$ and $A_{i+1}$ are diagonal in the same basis and the entries on
the diagonal can only be $\pm 1.$ Then, only the two cases outlined above
are possible.
\qed


{\it Alternative proof of Observation~1. }With the help of Lemma~5 one can consider each term of the KCBS
inequality and make there six possible replacements. For instance,
the term
$\mean{AB}$ may be replaced by $\mean{AB}\mapsto \pm \mean{B}$ (if one sets
$A\mapsto \pm \eins$) or $\mean{AB}\mapsto \pm \mean{A}$ (if one sets
$B\mapsto \pm \eins$) or $\mean{AB}\mapsto \pm 1.$ This results in a finite
set of $6^5=7776$ possible replacements. Some of them are contradictory and can be
disregarded, e.g., if one sets $B \mapsto \eins$ from the term $\mean{AB}$ and
$C\mapsto \eins$ from the term $\mean{CD}$, then one cannot set $\mean{BC}\mapsto - 1$
anymore. For the remaining replacements, one can directly check with a computer
that the $\mean{\chi_{\rm KCBS}}$ reduces to the classical bound.
\qed


\subsection*{A2: Detailed discussion of the generalized KCBS
inequalities}

First, we prove the following statement:

{\bf Lemma 6.} Consider the generalized KCBS operator
\be
\mean{\chi_{N}} = \sum_{i=1}^{N-1} \mean{A_i A_{i+1}} - \mean{A_N A_1}
\ee
for $N$ even, where the $A_i$ are dichotomic observables on a three-dimensional
system, which are not proportional to the identity. Furthermore, the commuting
pairs should not be equal, that is $A_i \neq A_{i+1}.$ Then, the bound
\be
\mean{\chi_{N}} \geq -(N-2)
\ee
holds.

{\it Proof of Lemma 6.} {From} the conditions, it follows that the observables
have to be of the form $A_i = \pm (\eins-2\ketbra{a_i})$ with
$\braket{a_i}{a_{i+1}}=0.$ This implies that the sequential measurements
can be rephrased via $A_i A_{i+1} = \pm (\eins - 2\ketbra{a_i}- 2\ketbra{a_{i+1}}).$
Let us first assume that the  signs in front of the $A_i$ are alternating,
that is, $A_i = + (\eins-2\ketbra{a_i})$ for odd $i$ and $A_i = - (\eins-2\ketbra{a_i})$
for even $i$. Then, a direct calculation leads to
\be
\mean{\chi_{N}} = -(N-2) + 4 \mean{\sum_{k=2}^{N-1}\ketbra{a_k}}.
\ee
{From} this, $\mean{\chi_{N}} \geq -(N-2)$ follows, since the operator in the sum
is positive semidefinite.

A general distribution of signs for the $A_i$ results in a certain distribution
of signs for the $A_i A_{i+1}$. If $I$ denotes the set of index pairs $(k,k+1)$,
where $A_k A_{k+1} = + (\eins - 2\ketbra{a_k}- 2\ketbra{a_{k+1}})$, then $I$ has
always an odd number of elements. We can then write:
\be
\mean{\chi_{N}} = -(N-2) + 2(|I|-1) + 4 \mean{\sum_{k=1}^{N}\alpha_k \ketbra{a_k}}
\ee
where $\alpha_k=1$ if both $(k,k+1)\notin I$ and $(k-1,k)\notin I$,
$\alpha_k=0$ if either $(k,k+1)\in I, (k-1,k)\notin I$
or $(k,k+1)\notin I, (k-1,k)\in I$, and
$\alpha_k=-1$ if both $(k,k+1)\in I$ and $(k-1,k)\in I$.

It remains to show that the last two terms are non-negative. The main
idea to prove this is to use the fact that an operator like
$X= \eins - \ketbra{a_i}-\ketbra{a_{i+1}}$ is positive semidefinite,
since $\ket{a_i}$ and $\ket{a_{i+1}}$ are orthogonal.

More explicitly, let us first consider the case where the index pairs
in $I$ are connected and distinguish different cases for the number of
elements in $I$. If $|I|=1$, there are no $k$ with $\alpha_k=-1$, so
$2(|I|-1) + 4 \mean{\sum_{k=1}^{N}\alpha_k \ketbra{a_k}} \geq 0.$
If $|I|=2$, then $I=\{(i-1,i),(i,i+1)\}$ and there is a single
$\alpha_i=-1$. In this case, one has
$2|I| + 4 \mean{\sum_{k=1}^{N}\alpha_k \ketbra{a_k}} \geq 0.$
This is not yet the desired bound, but it will be useful later.

If $|I|=3$, then $I=\{(i-1,i),(i,i+1),(i+1,i+2)\}$ and we have
$\alpha_i=\alpha_{i+1}=-1$. But now, the fact that
$X= \eins - \ketbra{a_i}-\ketbra{a_{i+1}} \geq 0$
directly implies that
$2(|I|-1) + 4 \mean{\sum_{k=1}^{N}\alpha_k \ketbra{a_k}} \geq 0.$
If $|I|=4$ there are three $\alpha_k =-1$ and we can use $X \geq 0$
two times, showing that again
$2|I| + 4 \mean{\sum_{k=1}^{N}\alpha_k \ketbra{a_k}} \geq 0.$
All this can be iterated, resulting in two different bounds,
for $|I|$ odd and $|I|$ even.

To complete the proof, we have to consider a general $I$ which does
not necessarily form a single block. One can then consider the different
blocks and,  since $|I|$ is odd, at least one of the blocks contains
an odd number of index pairs. Then, summing up the bound for the single
blocks leads to $2(|I|-1) + 4 \mean{\sum_{k=1}^{N}\alpha_k \ketbra{a_k}} \geq 0.$
\qed

Finally, in order to justify Eq.~(6) in the main text for the three-dimensional case,
we have to discuss what happens if one of the observables is proportional
to the identity. However, then the mean value $\mean{\chi_{N}}$ reduces to inequalities
which will be discussed later (see Lemma 9 in Appendix A4).

\subsection*{A3: Detailed proof of Observation~2}

For computing the minimal value in two-dimensional systems, we need the 
following Lemma. Note that the resulting value has been reported before
\cite{barbieri09}, so the main task is to prove rigorously that this is
indeed optimal.

{\bf Lemma~7. }Let $\ket{\af_i} \in \mathbb{R}^3$ be normalized
real three-dimensional vectors and define
\begin{subequations}
\begin{align}
\chi_N &= \sum_{i=1}^N \braket{\af_i}{\af_{i+1}} \mbox{ for $N$ odd,}
\\
\chi_N &= - \braket{\af_1}{\af_{2}} + \sum_{i=2}^N \braket{\af_i}{\af_{i+1}}
\mbox{ for $N$ even.}
\end{align}
\end{subequations}
Then we have
\be
\chi_N \geq - N \cos(\frac{\pi}{N}).
\label{cosbound}
\ee
{\it Proof of Lemma~7. }We write
$\ket{\af_i} = \{\cos(\alpha_i), \sin(\alpha_i) \cos(\beta_i),\sin(\alpha_i) \sin(\beta_i)\}$
and then we have
\begin{align}
\chi_{N} = & \sum_{i=1}^N [\pm]
\Big[\cos(\alpha_i)\cos(\alpha_{i+1})
\nonumber
\\
&+\cos(\beta_i-\beta_{i+1})\sin(\alpha_i)\sin(\alpha_{i+1})\Big],
\label{lemmaeq1}
\end{align}
where the symbol $[\pm]$ denotes the possibly changing sign of the term with $i=1$.
Let us first explain why the minimum of this expression can be obtained by setting all
the $\beta_i=0.$ Without losing generality, we can assume that $\ket{\af_1}$
points in the $x$-direction, i.e., $\alpha_1=0$ and $\sin(\alpha_1)=0.$ Then, only
$N-2$ terms of the type $\sin(\alpha_i)\sin(\alpha_{i+1})$ remain and all of them
have a positive prefactor. For given values of $\beta_i$ we can choose the signs
of $\alpha_2,\ldots, \alpha_{N-1}$ such that all these terms are negative, while
the other parts of the expression are not affected by this. Then, it is clearly
optimal to choose $\beta_2=\beta_3=\ldots =\beta_{N}=0.$ This means that all the
vectors lie in the $x$-$y$-plane.

Having set all $\beta_i=0$, the expression is simplified to
$\chi_N = \sum_{i=1}^N [\pm] \cos(\alpha_i-\alpha_{i+1}).$
We use the notation $\delta_i =\alpha_i-\alpha_{i+1}$ and minimize
$\sum_{i=1}^N [\pm] \cos(\delta_i)$ under the constraint
$\sum_{i=1}^{N}\delta_i=0.$ Using Lagrange multipliers, it follows that
$[\pm]\sin(\delta_i)=\lambda$ for all $i$.

If $N$ is odd, this means that we can express any $\delta_i$ as
$\delta_i=\pi/2 \pm \vartheta +2 \pi k_i$ with $\vartheta \geq 0.$
From $\cos(\pi/2 + \vartheta +2 \pi k_i) = - \cos(\pi/2 - \vartheta +2 \pi k_i),$
it follows that the sign in front of the $\vartheta$ should be identical for all
$\delta_i$, otherwise, the expression is not minimized. Let us first consider
the case that all signs a positive. From the condition $\sum_{i=1}^{N}\delta_i=0,$
it follows that $N (\pi/2) + N \vartheta + 2 \pi K = 0,$ with $K=\sum_{i=1}^N k_i.$
Since we wish to minimize $\chi_N,$ the angles $\delta_i$ should be as close as
possible to $\pi,$ which means that $|\vartheta-\pi/2|$ should be minimal. This leads
to the result that one has to choose $K = - (N \pm 1)/2.$ Computing the corresponding
$\vartheta$ leads to $\vartheta= \pi/2 \pm \pi/N,$ which results in Eq.~(\ref{cosbound}).
If the signs in front of all $\vartheta$ are negative, one can make a similar argument,
but this time has to minimize $|\delta_i + \pi|$ or $|\vartheta-3\pi/2|.$ This leads to
the same solutions.

If $N$ is even, one has for $i=2,\ldots,N$ again $\delta_i=\pi/2 \pm \vartheta +2 \pi k_i$
and the first $\delta_1$ can be written as
$\delta_1= - \pi/2 \pm \vartheta +2 \pi k_1$. One can directly see that if the
signs in front of $\vartheta$ is positive (negative) for all $i=2,\ldots,N$ it has to
be positive (negative) also for $i=1.$ A direct calculation as before leads to
$\vartheta = \pi/2 \pm \pi/N$ and, again, to the same bound of Eq.~(\ref{cosbound}).
\qed

{\it Proof of Observation~2. }Let us first assume that none of
the observables is proportional to the identity, and consider a
single sequential measurement $\mean{AB}$ of two dichotomic
noncommuting observables $A=\ketbra{A^+}-\ketbra{A^-}=P_+^A-P_-^A$ and
$B=\ketbra{B^+}-\ketbra{B^-}=P_+^B-P_-^B.$ We can also express $\ketbra{A^+}$
and $\ketbra{B^+}$ in terms of their Bloch vectors $\ket{\af}$ and
$\ket{\bfrak}.$ Then, we have that
\be
\mean{AB}=2 |\braket{A^+}{B^+}|^2-1 = \braket{\af}{\bfrak}.
\label{nicetrick}
\ee
Note that this means that the mean value $\mean{AB}$ is independent of the initial
quantum state. To see this relation, we write
$\mean{AB}
={\rm tr}(P^B_+P^A_+ \vr P^A_+P^B_+)
-
{\rm tr}(P^B_-P^A_+ \vr P^A_+P^B_-)
-
{\rm tr}(P^B_+P^A_- \vr P^A_-P^B_+)
+
{\rm tr}(P^B_-P^A_- \vr P^A_-P^B_-).
$
Using the fact that in a two-dimensional system
$|\braket{A^+}{B^+}|^2=|\braket{A^-}{B^-}|^2$
and
$|\braket{A^-}{B^+}|^2=|\braket{A^+}{B^-}|^2$ holds,
and ${\rm tr}(\vr)=1$, this can directly be simplified to the
above expression.
Using the above expression, we can write
$\mean{\chi_{\rm KCBS}}=\sum_{i=1}^5 \braket{\af_i}{\af_{i+1}}.$
Then, Lemma~7 proves the desired bound.

It remains to discuss the case where one or more observables in
the KCBS inequality are proportional to the identity. Let us first
assume that only one observable, say $A_1$ is proportional to the identity.
Then, if the Bloch vector of $\vr$ is denoted by $\ket{\mathfrak{r}}$
a direct calculation shows that the KCBS operator reads
\be
\mean{\chi_{\rm KCBS}} =
\braket{\mathfrak{r}}{\af_{2}}+
\sum_{i=2}^4 \braket{\af_i}{\af_{i+1}}
+
\braket{\af_5}{\mathfrak{r}},
\ee
and Lemma~7 proves again the claim. If two observables $A_i$
and $A_j$ are proportional to the identity, the same rewriting can be
applied, if $A_i$ and $A_j$ do not occur jointly in one correlation term.
This is the case if $j\neq i\pm 1.$ In the other case (say, $A_1=\eins$ and
$A_2=-\eins$), one has $\mean{A_1 A_2} = -1$ and can rewrite
\be
\mean{\chi_{\rm KCBS}} =
-1
-\braket{\mathfrak{r}}{\af_{2}}+
\sum_{i=3}^4 \braket{\af_i}{\af_{i+1}}
+ \braket{\af_4}{\mathfrak{r}},
\ee
and Lemma~7 implies that
$\mean{\chi_{\rm KCBS}} \geq - 4 \cos(\pi/4) - 1 = -2 \sqrt{2} -1
> - 5 \cos(\pi/5) = - 5 (1+\sqrt{5})/4.$ If more than two observables
are proportional to the identity, the bound can be proven similarly.
\qed


\subsection*{A4: Proof of Observation~3}


We need a whole sequence of Lemmata:


{\bf Lemma~8. }If one has three dichotomic measurements $A_i, i=1, 2, 3$
on a three-dimensional quantum system which commute pairwise
$[A_i, A_j]=0,$ then either
\begin{itemize}
\item[(a)] one of the observables is proportional to the identity,
$A_i = \pm \eins$ for some $i$ or,
\item[(b)] the product of two observables of the three observables is
proportional to the identity,
$A_i A_{j} = \pm \eins$ for some pair $i,j$ or,
\item[(c)] The product of all three observables is proportional to the identity,
$A_1 A_{1} A_3 = \pm \eins.$
\end{itemize}
Note that these cases are not exclusive and that for a triple of observables
several of these cases may apply at the same time.


{\it Proof of Lemma~8.} This can be proven in the same way as Lemma~5, since all
$A_i$ are diagonal in the same basis.
\qed


{\bf Lemma~9. }For sequences of dichotomic measurements the following
inequalities hold:
\be
\eta_N \equiv \mean{A_1}
+ \sum_{i=1}^{N-1}\mean{A_i A_{i+1}} - \mean{A_N} \leq N-1.
\ee
Here, it is always assumed that two observables which occur
in the same sequence commute. Moreover, if we define
\be
\zeta_N \equiv \sum_{i=1}^{N}\mean{A_i A_{i+1}} - \mean{A_N A_1},
\ee
then we have
\be
\zeta_N \leq N-2
\ee
in two-dimensional systems, while for three-dimensional systems.
\bea
\zeta_3 \leq 1; & & \zeta_4 \leq 2,
\\
\zeta_5 \leq \sqrt{5}(4-\sqrt{5}), & & \zeta_6 \leq 1+\sqrt{5}(4-\sqrt{5})
= 4(\sqrt{5}-1),
\nonumber
\eea
holds.


{\it Proof of Lemma~9. }If we consider $\eta_N$ for $N=2$ both observables
commute and the claim $\mean{A_1} +\mean{A_1A_2}-\mean{A_2} \leq 1$ is clear,
as it holds for any eigenvector. The bounds for general $\eta_N$ follow by
induction, where in each step of the induction
$\mean{A_N A_{N+1}}-\mean{A_{N+1}} \leq 1-\mean{A_N}$ is used, but this
is nothing but the bound for $N=2$.

The bounds for $\zeta_N$ are just the ones derived for the generalized KCBS
inequalities, see Eq.~(6) in the main text and Appendix A2.
$\qed$


{\bf Lemma~10.} Consider the PM square with dichotomic observables
on a three-dimensional system, where for one column and one row
only the case (c) in Lemma~8 applies. Then, one cannot violate
the classical bound and one has $\mean{\chi_{\rm PM}} \leq 4.$


{\it Proof of Lemma~10. }Let us consider the case that the condition holds
for the first column and the first row, the other cases are analogous. Then,
none of the observables $A, B, C, a, \alpha$ is proportional to the identity
since, otherwise, case (a) in Lemma~8 would apply. These observables
can all be written as
\be
A = \pm (\eins - 2 \ketbra{A}),
\ee
with some vector $\ket{A},$ and the vector $\ket{A}$ characterizes the observable
$A$ up to the total sign uniquely. In this notation, two observables $X$ and $Y$
commute if and only if the corresponding vectors $\ket{X}$ and $\ket{Y}$ are
the same or orthogonal. For our situation, it follows that the vectors $\ket{A}, \ket{B},$
and $\ket{C}$ form an orthonormal basis of the three-dimensional space, since if
two of them were the same, then for the first row also the case (b) in Lemma~8 would apply. Similarly, the vectors $\ket{A}, \ket{a}$ and $\ket{\alpha}$
form another orthonormal basis of the three-dimensional space. We can distinguish
two cases:

{\it Case 1: The vector $\ket{B}$ is neither orthogonal nor parallel to $\ket{a}.$}
{From} this, it follows that $\ket{B}$ is also neither orthogonal nor parallel
to $\ket{\alpha}$ and similarly, $\ket{C}$ is neither orthogonal nor parallel
to $\ket{a}$ and $\ket{\alpha}$ and vice versa.

Let us consider the observable $b$ in the PM square. This observable can be proportional
to the identity, but if this is not the case, the corresponding vector $\ket{b}$ has to
be parallel or orthogonal to $\ket{B}$ and $\ket{a}.$ Since $\ket{B}$ and $\ket{a}$ are
neither orthogonal nor parallel, it has to be orthogonal to both, which means that it is
parallel to $\ket{A}.$ Consequently, the observable $b$ is either proportional to the
identity or proportional to $A$. Similarly, all the other observables $\beta, c,$ and
$\gamma$ are either proportional to the identity or proportional to $A$.

Let us now consider the expectation value of the PM operator $\mean{\chi_{\rm PM}}$ for
some quantum state $\vr$. We denote this expectation value as $\mean{\chi_{\rm PM}}_{\vr}$
in order to stress the dependence on $\vr.$ The observable $A$ can be written as
$A=P_+ -P_-,$ where $P_+$ and $P_-$ are the projectors onto the positive or negative
eigenspace. One of these projectors is one-dimensional and equals $\ketbra{A},$ the other
other one is two-dimensional. For definiteness, let us take $P_+ = \ketbra{A}$ and
$P_-=\eins-\ketbra{A}.$

Instead of $\vr$, we may consider the depolarized state $\sigma= p_+ \vr_+ + p_- \vr_-$, with
$\vr_\pm= P_\pm \vr P_\pm / p_\pm$ and $p_\pm={\rm tr}(P_\pm \vr P_\pm).$ Our first claim is that,
in our situation,
\be
\mean{\chi_{\rm PM}}_{\vr} {=} \mean{\chi_{\rm PM}}_{\sigma}
=p_+\mean{\chi_{\rm PM}}_{\vr_+}+p_-\mean{\chi_{\rm PM}}_{\vr_-}.
\label{zwischenresultat1}
\ee
It suffices to prove this for all rows and columns separately. Since the observables
in each column or row commute, we can first measure observables which might be
proportional to $A$. For the first column and the first row the statement
is clear: We first measure $A$ and the result is the same for $\vr$ and $\sigma.$
After the measurement of $A$, however, the state $\vr$ is projected either
onto $\vr_+$ or $\vr_-.$ Therefore, for the following measurements it does not matter
whether the initial state was $\vr$ or $\sigma.$ As an example for the other
rows and columns, we consider the second column. Here, we can first measure
$\beta$ and then $b$ and finally $B.$ If $\beta$ or $b$ are proportional to
$A$, then the statement is again clear. If both $\beta$ and $b$ are proportional
to the identity, then the measurement of $\mean{\beta b B}_\vr$ equals
$\pm \mean{B}_\vr.$ Then, however, one can directly calculate that
$\mean{B}_\vr=\mean{B}_\sigma,$ since $B$ and $A$ commute.

Having established the validity of Eq.~(\ref{zwischenresultat1}), we proceed by showing that
for for each term $\mean{\chi_{\rm PM}}_{\vr_+}$ and $\mean{\chi_{\rm PM}}_{\vr_-}$ separately
the classical bound holds. For $\mean{\chi_{\rm PM}}_{\vr_+}$ this is clear: Since $P_+=\ketbra{A}$,
we have that $\vr_+=\ketbra{A}$ and $\ket{A}$ is an eigenvector of all observables occurring in
the PM square. Therefore, the results obtained in $\mean{\chi_{\rm PM}}_{\vr_+}$ correspond to a
classical assignment of $\pm 1$ to all observables, and $\mean{\chi_{\rm PM}}_{\vr_+}\leq 4$ follows.
For the other term $\mean{\chi_{\rm PM}}_{\vr_-}$, the problem is effectively a two-dimensional one,
and we can consider the restriction of the observables to the two-dimensional space, e.g.,
$\bar{A}= P_- A P_-$, etc. In this restricted space we have that $\bar{A}, \bar{b}, \bar{\beta},
\bar{c},$ and $\bar{\gamma}$ are all of them proportional to the identity and, therefore, result in a classical
assignment $\pm 1$ independent of $\vr_-.$ Let us denote these assignments by
$\hat{A}, \hat{b}, \hat{\beta}, \hat{c},$ and $\hat{\gamma}.$ Then, it remains to be shown that
\bea
\ZZ&=&\hat{A} \big[\mean{\bar{B}\bar{C}}_{\vr_-} + \mean{\bar{\alpha}\bar{a}}_{\vr_-}\big]
+ \hat{b}\hat{c}\mean{\bar{a}}_{\vr_-}
\nonumber
\\
&& + \hat{\beta}\hat{\gamma}\mean{\bar{\alpha}}_{\vr_-}
+ \hat{b}\hat{\beta}\mean{\bar{B}}_{\vr_-}
- \hat{c}\hat{\gamma}\mean{\bar{C}}_{\vr_-}
\leq 4
\label{lemma7eq3}
\eea
for all classical assignments and for all states $\vr_-.$ For observables
$\bar{B}$ and $\bar{C}$ we have furthermore that $\bar{B}\bar{C}=\pm \eins$
(see Lemma~5), hence $\bar{B}= \pm \bar{C}$ and similarly
$\bar{a}= \pm \bar{\alpha}.$ If one wishes to maximize $\ZZ$ for the case
$\hat{A}=+1$, one has to choose $\bar{B} =\bar{C}$ and $\bar{a}= \bar{\alpha}.$
Then, the product of the four last terms in $\ZZ$ equals $-1,$ and $\ZZ \leq 4$
holds. For the case $\hat{A}=-1$ one chooses $\bar{B} = -\bar{C}$ and $\bar{a}= -\bar{\alpha},$
but still the product of the four last terms in $\ZZ$ equals $-1,$ and $\ZZ \leq 4.$
This finishes the proof of the first case.

{\it Case 2: The bases $\ket{A}, \ket{B}, \ket{C}$ and $\ket{A}, \ket{a}, \ket{\alpha}$ are
(up to some permutations or signs) the same.} For instance, we can have the case
in which $\ket{B}=\ket{a}$ and $\ket{C}=\ket{\alpha};$ the other possibilities can be treated
similarly.

In this case, since $\ket{B}$ and $\ket{\alpha}$ are orthogonal, the observable $\beta$
has to be either proportional to the identity or proportional to $A$. For the same
reason, $c$ has to be either proportional to the identity or to $A$.

Let us first consider the case in which one of the observables $\beta$ and $c$ is proportional
to $A$, say $\beta=\pm A$ for definiteness. Then, since $\ket{\beta}=\ket{A}$ and
$\ket{B}$ are orthogonal, $b$ can only be the identity or proportional to $C.$ Similarly,
$\gamma$ can only be the identity or proportional to $C.$ It follows that {\it all} nine
observables in the PM square are diagonal in the basis $\ket{A}, \ket{B}, \ket{C}$, and all
observables commute. Then, $\mean{\chi_{\rm PM}}\leq 4$ follows, as this inequality holds
in any eigenspace.

Second, let us consider the case in which $\beta$ and $c$ are both proportional to the identity.
This results in fixed assignments $\hat{\beta}$ and $\hat{c}$ for them. Moreover, $B$ and $a$
differ only by a sign $\hat{\mu}$ (that is, $a=\hat{\mu}B$) and $C$ and $\alpha$ differ only by a
sign $\hat{\nu}$ (i.e., $\alpha=\hat{\nu}C$). So we have to consider
\bea
\XX &=&
\mean{ABC}+\hat{\mu}\hat{\nu}\mean{ABC} + \hat{\beta}\mean{Bb}
\nonumber
\\
&&+\hat{\mu}\hat{c}\mean{Bb}
+\hat{\nu}\hat{\beta}\mean{C\gamma} - \hat{c}\mean{C\gamma}.
\eea
In order to achieve $\XX>4$ one has to choose $\hat{\mu}=\hat{\nu},$
$\hat{\beta}=\hat{\mu}\hat{c},$ and $\hat{c}=-\hat{\nu}\hat{\beta}.$
However, the later is equivalent to $\hat{\beta}=-\hat{\nu}\hat{c},$ showing
that this assignment is not possible. Therefore, $\XX \leq 4$ has to hold.
This finishes the proof of the second case.
$\qed$


{\bf Lemma~11.} Consider the PM square with dichotomic observables
on a three-dimensional system, where for one column (or one row)
only the case (c) in Lemma~8 applies. Then, one cannot violate
the classical bound and one has $\mean{\chi_{\rm PM}} \leq 4.$


{\it Proof of Lemma~11.} We assume that the condition holds
for the first column. Then, none of the observables $A, a,$ and $\alpha$
are proportional to the identity, and the corresponding vectors
$\ket{A}, \ket{a},$ and $\ket{\alpha}$ form an orthonormal basis
of the three-dimensional space.

The idea of our proof is to consider possible other observables
in the PM square, which are not proportional to the identity, but also
not proportional to $A, a,$ or $\alpha.$ We will see that there are
not many possibilities for the observables, and in all cases the bound
$\mean{\chi_{\rm PM}} \leq 4$ can be proved explicitly.

First, consider the case that there {\it all} nontrivial
observables in the PM square are proportional to $A, a,$
or $\alpha.$ This means that all observables in the PM square are
diagonal in the basis defined by $\ket{A}, \ket{a},$ and $\ket{\alpha},$
and all observables commute. But then the bound $\mean{\chi_{\rm PM}} \leq 4$
is clear.

Second, consider the case that there are several nontrivial observables,
which are {\it not} proportional to $A, a,$ or $\alpha.$
Without losing
generality, we can assume that the first of these observables is $B.$
This implies that $\ket{B}$ is orthogonal to $\ket{A}$ and lies in the
plane spanned by $\ket{a}$ and $\ket{\alpha},$ but
$\ket{a} \neq \ket{B} \neq \ket{\alpha}.$ It follows for the observables $b$
and $\beta$ that they can only be proportional to the identity or to $A$
(see Case~1 in Lemma~10). We denote this as $b=\hat{b}[A],$ where
$[A]=A$ or $\eins,$ and $\hat{b}$ denotes the proper sign, i.e., $b=\hat{b}A$
or $b=\hat{b}\eins.$ Similarly, we write $\beta=\hat{\beta}[A].$

Let us assume that there is a second nontrivial observable which is not
proportional to $A, a,$ or $\alpha$ (but it might be proportional to $B$).
We can distinguish three cases:

(i) First, this observable can be given by $C$ and $C$ is not proportional to
$B.$ Then, this is exactly the situation of Case 1 in Lemma~10, and
$\mean{\chi_{\rm PM}} \leq 4$ follows.

(ii) Second, this observable can be given by $C$. However, $C$ is proportional
to $B.$ Then, $c=\hat{c}[A]$ and $\gamma=\hat{\gamma}[A]$ follows.
Now the proof can proceed as in Case 1 of Lemma~10. One arrives to the same
Eq.~(\ref{lemma7eq3}), with the extra condition that $\bar{B}= \pm \bar{C}$, which
was deduced after Eq.~(\ref{lemma7eq3}) anyway. Therefore, $\mean{\chi_{\rm PM}} \leq 4$
has to hold.

(iii) Third, this observable can be given by $c.$ Then, it cannot be proportional
to $B,$ since $\ket{B}$ is not orthogonal to $\ket{a}.$ It first follows that
$C=\hat{C}[a]$ and $\gamma=\hat{\gamma}[a].$ Combined with the properties
of $B,$ one finds that $C=\hat{C}\eins$ and $b=\hat{b}\eins$ has to hold. Then,
the PM inequality reads
\bea
\mathcal{Y}&=&
\mean{A\alpha a}+ \mean{B(A\hat{C}+\hat{b}\hat{\beta}[A])}
\nonumber
\\
&& +\hat{\beta}\hat{\gamma}
\mean{\alpha[A][a]}
+\mean{c(\hat{b}a-\hat{C}\hat{\gamma}[a])}.
\label{lemma8eq1}
\eea
In this expression, the observables $B$ and $c$ occur only in a single term and a single
context. Therefore, for any quantum state, we can obtain an upper bound on $\mathcal{Y}$
by replacing $B\mapsto \pm \eins$ and $c\mapsto \pm \eins$ with appropriately chosen signs.
However, with this replacement, all observables occurring in $\mathcal{Y}$ are diagonal in the
basis defined by $\ket{A}, \ket{a},$ and $\ket{\alpha},$
and $\mathcal{Y}=\mean{\chi_{\rm PM}} \leq 4$ follows.

In summary, the discussion of the cases (i), (ii), and (iii) has shown the following:
It is not possible to have three nontrivial observables in the PM square, which are all of them
not proportional to $A, a,$ or $\alpha.$ If one has two of such observables, then the classical
bound has been proven.

It remains to be discussed what happen if one has only one observable (say, $B$), which
is not proportional to $A, a,$ or $\alpha.$ However, then the PM inequality can be written
similarly as in Eq.~(\ref{lemma8eq1}), and $B$ occurs in a single context. We can
set again $B\mapsto \pm \eins$ and the claim follows.
$\qed$

Finally, we can prove our Observation~3:


{\it Proof of Observation~3. }Lemma~10 and Lemma~11 solve the problem, if case (c)
in one column or row happens. Therefore, we can assume that in all columns and all rows
only the cases (a) or (b) from Lemma~8 apply. However, in these cases, we obtain a simple
replacement rule: For case (a), one of the observables has to be replaced with a
classical value $\pm 1$ and, for case (b), one of the observables can be replaced by
a different one from the same row or column. In both cases, the PM inequality
is simplified.

For case (a), there are six possible replacement rules, as one of the three observables
must be replaced by $\pm 1$. Similarly, for case (b), there are six replacement rules.
Therefore, one obtains a finite number, namely $(6+6)^6$ possible replacements. As in the
case of the KCBS inequality (see the alternative proof of Observation~1 in Appendix~A1),
some of them lead to
contradictions (e.g., one may try to set $A=+ \eins$ from the first column, but
$A= -\eins$ holds due to the rule from the first row). Taking this into account,
one can perform an exhaustive search of all possibilities, preferably by
computer. For all cases, either the classical bound holds trivially (e.g., because
the assignments require already, that one row is $-1$) or the PM inequality can
be reduced, up to some constant, to one of the inequalities in Lemma~8. In most cases,
one obtains the classical bound. However, in some cases, the PM inequality is reduced to
$\mean{\chi_{\rm PM}} = \zeta_5+1$ or $\mean{\chi_{\rm PM}} = \zeta_6.$ To give an example,
one may consider the square
\be
\begin{bmatrix}
A&B&C\\
a& b& c\\
\alpha &\beta & \gamma
\end{bmatrix}
=
\begin{bmatrix}
A&\eins&C\\
a& b& \eins\\
\eins &\beta & \gamma
\end{bmatrix},
\ee
which results in $\mean{\chi_{\rm PM}} = \zeta_6$ for appropriately
chosen $A_i.$ Therefore, from Lemma~9 follows that in three
dimensions $\mean{\chi_{\rm PM}}=4(\sqrt{5}-1) \approx 4.94$ holds
and can indeed be reached.
$\qed$


\subsection*{A5: Imperfect measurements}


In this section we discuss the noise robustness of Observation 4. In the first
subsection, we prove that Observation 4 also holds for the model of noisy
measurements explained in the main text. In the second subsection, we discuss
a noise model that reproduces the probabilities of the most general POVM.


\subsubsection*{\bf A5.1: Noisy measurements}


In order to explain the probabilities from a noisy measurement, we first
consider the following measurement model: Instead of performing the
projective measurement $A$, one of two possible actions are taken:
\begin{itemize}
\item[(a)] with a probability $p_A$ the projective measurement is
performed, or
\item[(b)] with a probability $1-p_A$ a completely random outcome $\pm 1$
is assigned independently of the initial state. Here, the results $+1$ and $-1$
occur with equal probability.
\end{itemize}
In case (b), after the assignment the physical system is
left in one of two possible states $\vr^+$ or $\vr^-$, depending
on the assignment. We will not make any assumptions on
$\vr^\pm$.

Before formulating and proving a bound on $\mean{{\chi}_{\rm PM}}$ in this
scenario, it is useful to discuss the structure of $\mean{{\chi}_{\rm PM}}$
for the measurement model. A single measurement sequence $\mean{ABC}$ is
split into eight terms: With a prefactor $p_A p_B p_C$ one has the
value, which is obtained, if all measurements are projective; with
a prefactor $p_A p_B (1-p_C)$ one has the value, where $A$ and $B$ are
projective, and $C$ is a random assignment, etc. It follows that the
total mean value $\mean{{\chi}_{\rm PM}}$ is an affine function in the
probability $p_A$ (if all other parameters are fixed) and also in all
other probabilities $p_X$ for the other measurements. Consequently, the
maximum of $\mean{{\chi}_{\rm PM}}$ is attained either at $p_A=1$ or
$p_A=0,$ and similarly for all the measurements. Therefore, for maximizing
$\mean{{\chi}_{\rm PM}}$ it suffices to consider the finite set of cases
where, for each observable, either always possibility (a) or always
possibility (b) is taken. We can formulate:

{\bf Proposition 12.} Consider noisy measurements as described above.
Then, the bound from Observation 4
\be
\mean{{\chi}_{\rm PM}} \leq 3\sqrt{3}
\ee
holds.

{\it Proof.}
As discussed above, we only have to discuss a finite number of
cases. Let us consider a single term $\mean{ABC}$. If $C$ is a
random assignment, then $\mean{ABC} = 0$, independently how $A$
and $B$ are realized. It follows that if $C, \beta$ or $a$ are
random assignments, then $\mean{{\chi}_{\rm PM}} \leq 4$.

On the other hand, if $A$ is a random assignment, then
$\mean{ABC} = 0$ as well: (i) If $B$ and $C$ are projective,
then the measurement of $B$ and $C$ results in the state
independent mean value $\mean{BC}$ [see Eq.~(8) in the
main text]. This value is independent of the state
$\vr^\pm$ remaining after the assignment of $A$, hence
$\mean{ABC} = \mean{AB}-\mean{AB}=0.$ (ii) If $B$ is a
random assignment, one can also directly calculate that
$\mean{ABC} = 0$ and the case that (iii) $C$ is a random
assignment has been discussed already. Consequently, if
$A, b,$ or $\gamma$ are random assignments, then
$\mean{{\chi}_{\rm PM}} \leq 4$.

It remains to discuss the case that $B, c,$ or $\alpha$ are
random assignments while all other measurements are
projective. First, one can directly calculate that if $A,C$
are projective, and $B$ is a random assignment, then
\be
\mean{ABC}=\trace(\vr A) \trace(C X),
\label{prop12a}
\ee
with $X=(\vr^+-\vr^-)/2.$ If $X$ is expressed in terms of
Pauli matrices, then the length of its Bloch vector does not
exceed one, since the Bloch vectors of $\vr^\pm$ are subnormalized.

The estimate of $\mean{{\chi}_{\rm PM}}$ can now proceed as in
the proof of Observation 4, and one arrives at the situation
of Lemma 7 in Appendix A3, where now the vectors are subnormalized,
and not necessarily normalized. But still the bound from Lemma 7
is valid: If the smallest vector in $\chi_6$ has a length $\omega,$
one can directly see that $\chi_6 \geq \omega[-N \cos(\pi/N)]-(1-\omega)4.$
This proves Proposition 12.
\qed


\subsubsection*{\bf A5.2: More general POVMs}


Now we consider a general dichotomic positive operator valued measure
(POVM) on a qubit system. This is characterized by two effects $E^+$ and
$E^-$, where $E^+ + E^- = \eins$ and the probabilities of the measurement
results are $p^+={\rm tr}(\vr E^+)$ and $p^-={\rm tr}(\vr E^-).$

These effects have to commute and one can write
$E^+=\alpha \ketbra{0}+\beta\ketbra{1}$ and
$E^-=\gamma \ketbra{0}+\delta\ketbra{1}$ in
an appropriate basis. We can assume that $\alpha \geq \beta$
and consequently $\delta \geq \gamma.$ Furthermore, it is no restriction
to choose $\beta \leq \gamma.$ Then, the effects can be written as
$E^+=\beta \eins+(\alpha-\beta)\ketbra{0}$ and
$E^-=\beta \eins + (\gamma - \beta)\eins+(\alpha - \beta)\ketbra{1}.$
This means that one can interpret the probabilities of the POVM as coming from
the following procedure: With a probability of $2\beta$ one assigns a random
outcome, with a probability of $\gamma -\beta$ one assigns the fixed value
$-1$, and with a probability of $(\alpha - \beta)$ one performs the projective
measurement.

This motivates the following measurement model: Instead of performing the
projective measurement $A$, one of three possible actions are taken:
\begin{itemize}
\item[(i)] with a probability $p_1^A$ the projective measurement is performed, or
\item[(ii)] with a probability $p_2^A$ a fixed outcome $\pm 1$ is assigned
independently of the initial state. After this announcement, the state is
left in the corresponding eigenstate of $A$, or
\item[(iii)] with a probability $p_3^A$ a completely random outcome $\pm 1$ is
assigned independently of the initial state.
\end{itemize}
As above, in case (iii), the physical system is left in one of two possible
states $\vr^+$ or $\vr^-$, but we will not make any assumptions on $\vr^\pm$.
For this measurement model, we have:

{\bf Proposition 13.} In the noise model described above, the PM operator is
bounded by
\be
\mean{{\chi}_{\rm PM}}\leq 1+\sqrt{9+6\sqrt{3}}\approx 5.404.
\ee

{\it Proof.} As in the proof of Proposition 12, we only have to consider a
finite set of cases. Let us first discuss the situation, where for each
measurement only the possibilities (i) and (ii) are taken.

First, we have to derive some formulas for sequential measurements. The reason
is that, if the option (ii) is chosen, then the original formula for sequential
measurements, Eq.~(15) in the main text, is not appropriate anymore
and different formulas have to be used.

In the following, we write $A= (\pm)_A$ if $A$ is a fixed assignment as described
in possibility (ii) above. If not explicitly stated otherwise, the observables
are measured as projective measurements. Then one can directly calculate that
\begin{subequations}
\bea
\mean{ABC} &=& (\pm)_A \mean{BC} \mbox{ if } A=(\pm)_A,
\label{testgl1}
\\
\mean{ABC} &=& \trace(\vr A) \mean{BC} \mbox{ if } B=(\pm)_B,
\label{testgl2}
\\
\mean{ABC} &=& (\pm)_C \mean{AB} \mbox{ if } C=(\pm)_C,
\label{testgl3}
\eea
\end{subequations}
Note that in Eq.~(\ref{testgl2}) there is no deviation from the
usual formula Eq.~(15) in the main text. Furthermore, we have
\begin{subequations}
\bea
\mean{ABC} &= &(\pm)_A (\pm)_B \trace(C \ketbra{B^\pm})
=(\pm)_A \mean{BC}
\nonumber
\\
&&\mbox{ if } A=(\pm)_A \mbox{ and } B=(\pm)_B,
\label{hilfgl1}
\\
\mean{ABC} &=& (\pm)_A (\pm)_C \trace(B \ketbra{A^\pm})
=(\pm)_C \mean{AB}
\nonumber
\\
&&\mbox{ if } A=(\pm)_A \mbox{ and } C=(\pm)_C,
\label{hilfgl2}
\\
\mean{ABC} &= &(\pm)_B (\pm)_C \trace(\vr A)
\nonumber
\\
&&
\mbox{ if } B=(\pm)_B \mbox{ and } C=(\pm)_C.
\label{hilfgl3}
\eea
\end{subequations}
In Eqs.~(\ref{hilfgl1}) and (\ref{hilfgl2}), $\ket{B^\pm}$ and $\ket{A^\pm}$
denote the eigenstates of $B$ and $A$, which are left after the fixed
assignment.

Equipped with these rules, we can discuss the different cases. First, from
Eqs.~(\ref{testgl1}), (\ref{testgl2}), and (\ref{hilfgl1}) it follows that
the proof of Observation 4 does not change, if fixed assignments
are made only on the observables which are measured at first or
second position of a sequence (i.e., the observables
$A, b, \gamma, B, c,$ and $\alpha$).

However, the structure of the inequality changes if one of the last measurements is a fixed assignment. 
To give an example, consider the
case that the measurement $\beta$ is a fixed assignment [case (ii) above],
while all other measurements are projective [case (i) above].
Using Eq.~(\ref{testgl3}) we have to estimate
\begin{align}
\XX & =
\mean{A}\mean{BC}+\mean{A}\mean{\alpha a}+
\mean{b}\mean{ca}
\nonumber
\\
&+\mean{b B}(\pm)_\beta
+\mean{\gamma \alpha}(\pm)_\beta
-\mean{\gamma}\mean{cC}.
\label{xeq1}
\end{align}
On can directly see that it suffices to estimate
\begin{align}
\XX' & =
\mean{B|C}+\mean{\alpha|a}+
\mean{\vr|b}\mean{c|a}
\nonumber
\\
&+\mean{b|B}
+\mean{\gamma|\alpha}
-\mean{\vr|\gamma}\mean{c|C},
\label{xeq2}
\end{align}
where all expressions should be understood as scalar products of the
corresponding Bloch vectors.
Then, a direct optimization over the three-dimensional Bloch vectors
proves that here
\be
\XX' \leq 1+\sqrt{9+6\sqrt{3}} \approx 5.404
\label{xeq3}
\ee
holds. In general, the observables $\beta, C,$ or $a$ are the possible third
measurements in a sequence. One can directly check that, if one or several of them are
fixed assignments, then an expression analogue to Eq.~(\ref{xeq1}) arises and
the bound of Eq.~(\ref{xeq3}) holds. Finally, if some of the $\beta, C,$ or $a$
are fixed assignments and, in addition, some of the $A, b, \gamma, B, c,$ and
$\alpha$ are fixed assignments, then the comparison between Eq.~(\ref{testgl3})
and Eqs.~(\ref{hilfgl2}) and (\ref{hilfgl3}) shows that no novel types of
expressions occur.

It remains to discuss the case where not only the possibilities (i) and
(ii) occur, but for one or more measurements also a random assignment
[possibility (iii)] is realized. As in the proof of Proposition 12, one
finds that only the cases where the second measurements ($B, c,$ and
$\alpha$) are random are interesting. In addition to Eq.~(\ref{prop12a})
one finds that $\mean{ABC}= (\pm_A) \trace(C X)$ if $B$ is random and $A$ is
a fixed assignment, and $\mean{ABC}= 0$ if $B$ is random and $C$ is
a fixed assignment. This shows that no new expressions occur, and
proves the claim.
\qed

Finally, we would like to add two remarks. First, it should be stressed that
the presented noise model still makes assumptions about the measurement,
especially about the post measurement state. Therefore, it is not the most
general measurement, and we do not claim that the resulting dimension
witnesses are device-independent.

Second, we would like to emphasize that the chosen order of the
measurements in the definition in Eq.~(9) in the main
text is important for the proof of the bounds for noisy measurements:
For other orders, it is not clear whether the dimension witnesses are
robust against imperfections. In fact, for some choices one finds that
the resulting inequalities are {\it not} robust against imperfections:
Consider, for instance, a measurement order, where one observable (say,
$\gamma$ for definiteness) is the second observable in one context and
the third observable in the other context. Furthermore, assume that $\gamma$
is an assignment [case (iii) above], while all other measurements are projective.
Then, we have to use Eq.~(\ref{testgl2}) for the first context of $\gamma$,
and Eq.~(\ref{testgl3}) for the second context. In Eq.~(\ref{testgl2}) there
is no difference to the usual formula, especially the formula does not
depend on the value assigned to $\gamma$. Eq.~(\ref{testgl3}), however,
depends on this value. This means that, for one term in the PM inequality, the
sign can be changed arbitrarily and so $\mean{{\chi}_{\rm PM}}=6$ can be reached.

\end{document}